\documentclass{article}

\usepackage{arxiv}

\usepackage{multirow}
\usepackage{graphicx}% Include figure files
\usepackage{dcolumn}% Align table columns on decimal point
\usepackage{bm}
\usepackage[mathlines]{lineno}% Enable numbering of text and display math

\newcommand{\bs}{\boldsymbol}

\usepackage{hyperref}
\hypersetup{
    colorlinks=true,
    linkcolor=blue,
    citecolor=blue,
    filecolor=magenta,      
    urlcolor=blue,
    hypertexnames=true
    }

\usepackage{natbib}
\usepackage{epstopdf, epsfig}
\usepackage{amssymb}
\usepackage{amsmath}
\usepackage[nameinlink, noabbrev]{cleveref}    
\usepackage{xcolor}
\usepackage{subfig}
\usepackage{float}
\usepackage[normalem]{ulem}
\usepackage{soul}
\usepackage{colortbl,xcolor}

\usepackage{booktabs}       % professional-quality tables
\usepackage{nicefrac}       % compact symbols for 1/2, etc.
\usepackage{microtype}      % microtypography

\title{Data-driven oscillator model for turbulent flows with multiple dominant frequencies}

\author{Youngjae Kim$^1$\thanks{Corresponding author: youngjaekim@ucla.edu} ,
        Koichiro Yawata$^2$,
        Hiroya Nakao$^{2,3}$,
        Kunihiko Taira$^1$ \\
        $^1$Department of Mechanical and Aerospace Engineering, University of California, Los Angeles, CA 90095, USA \\
        $^2$Department of Systems and Control Engineering, Institute of Science Tokyo, Tokyo 152-8552, Japan \\
        $^3$Research Center for Autonomous Systems Materialogy, Institute of Science Tokyo, Tokyo 152-8552, Japan
        }

\begin{document}

\maketitle

\begin{abstract}

The complex dynamics of high-dimensional oscillatory flows can be simplified using phase-reduction analysis, providing a deeper understanding of the flow response to external perturbations. Although phase-based modeling and analysis have been utilized in recent studies on oscillatory fluid flows, their usage is often limited to single-frequency flows due to difficulties in addressing chaotic characteristics, emerging from multiple dominant frequencies with broadband spectra in turbulent flows. In order to overcome this limitation, we propose a data-driven framework that models the dynamics of turbulent flows with multiple dominant frequencies based on a set of oscillators. The representative oscillators are extracted from the flow field data by training specially designed autoencoders. The oscillator dynamics are modeled through a machine-learning technique using neural networks to accurately predict the multi-frequency oscillatory behavior of turbulent flows. We verify the oscillator-based model of the oscillatory turbulent flows by applying the proposed data-driven method to the three-dimensional supersonic turbulent flow over a cavity. We show that the oscillators extracted from the spanwise-averaged pressure field represent the dominant large-scale flow features and reflect the physical characteristics of the turbulent cavity flow. The data-driven oscillator dynamics model with observation-based correction accurately forecasts the oscillatory behavior of the turbulent cavity flow for a long period. The proposed data-driven method for reduced-order modeling of turbulent flows with oscillators will enable deeper investigations of perturbation dynamics and control of turbulent flows.

\end{abstract}

\section{Introduction}    \label{sec:1}

Oscillations are prevalent in various dynamical systems in nature, including structural and mechanical vibrations, chemical and electrical oscillations, circadian and cardiac rhythms \citep{Pikovsky_Rosenblum_Kurths_2001, kuramoto2003chemical, winfree1967biological, winfree1980geometry}. Unsteady fluid flows are also prominent examples that exhibit oscillatory flow patterns, which contribute to their complex nonlinear dynamics. Such oscillatory behavior of flows can exert beneficial or detrimental effects on the performance and operation of systems in various engineering applications. Oscillations play an important role in structural damage due to flow-induced vibrations \citep{sarpkaya2004critical, williamson2004vortex}, noise generation \citep{inoue2002sound,buhler2014laminar}, mixing characteristics \citep{celik2009mixing,handa2014supersonic}, and efficiency of energy harvesting devices \citep{zhang2017design, wang2020state}. Therefore, analysis of oscillatory flows is essential in extending our understanding of the underlying flow physics and improving the design and reliability of engineering systems.

For the systematic investigation of the oscillatory behavior of fluid flows, considerable efforts have been devoted to developing various data-driven and operator-based modal analysis techniques \citep{taira2017modal, taira2020modal}. Proper orthogonal decomposition (POD) is one of the most popular modal analysis techniques, linearly decomposing the flow into spatially orthogonal modes that capture energetic components in the flow \citep{lumley1970,sirovich1987turbulence,berkooz1993proper,holmes2012turbulence}. POD has been further extended to flow analysis in the frequency domain, referred to as spectral proper orthogonal decomposition (SPOD), which identifies dominant frequencies along with their corresponding modal structures in unsteady flows \citep{towne2018spectral, nekkanti2021frequency}. Dynamic mode decomposition (DMD) also captures modal structures associated with characteristic frequencies and growth rates of the flow structure by the eigendecomposition of the linear evolution operator inferred from the flow field data \citep{schmid2010dynamic, schmid2022dynamic}. On the theoretical side, global linear stability analysis examines the eigenvalue spectra of the linearized Navier-Stokes operator, identifying unstable modes which trigger unsteadiness of the flow and transition to turbulence \citep{theofilis2011global}. Resolvent analysis supplements the global stability analysis by considering the input-output mechanisms of the flow \citep{trefethen1993hydrodynamic,jovanovic2005componentwise,mckeon2010critical,rolandi2024invitation}.

Complementary to the modal analysis techniques in understanding unsteady flows, phase-reduction analysis offers distinctive physical insights into their perturbation dynamics. Phase-reduction analysis simplifies the description of dynamical systems by projecting them onto an oscillator \citep{nakao2016phase}. Phase-reduction analysis facilitates theoretical investigations of synchronization between periodic forcing and system dynamics by characterizing the phase dependence of the system response to external perturbations. It has been applied to analyzing nonlinear oscillators in various fields, including biological rhythms \citep{winfree1967biological, smeal2010phase, kralemann2013vivo} and chemical oscillators \citep{kuramoto2003chemical, bomela2018optimal}.

In recent studies, phase-reduction analysis has been utilized to investigate unsteady flows. Various studies have applied phase-reduction analysis to investigate the wake synchronization characteristics of a two-dimensional flow around a bluff body, such as a circular cylinder \citep{taira2018phase, khodkar2020phase, kawamura2022adjoint} and an airfoil \citep{kawamura2022adjoint, godavarthi2023optimal}, with respect to local momentum injection, identifying the optimal forcing direction and locations for synchronization. Phase reduction analysis can also predict lock-on phenomena in fluid-structure interaction problems, such as wake synchronizations to the forced vibration of the cylinder \citep{khodkar2021phase} and surrounding elastic structures \citep{loe2021phase}. More recently, thermoacoustic oscillations in a Rijke tube have been analyzed using phase-reduction analysis \citep{skene2022phase}. Beyond the theoretical investigation of oscillatory flows, the phase-reduction technique has also guided the establishment of control schemes for unsteady oscillatory flows. For instance, optimal control schemes have been developed using the phase-reduction technique for rapid adjustment of the flow state \citep{nair2021phase}, fast wake synchronization to the periodic forcing \citep{godavarthi2023optimal}, and attenuation of transient lift fluctuations of an airfoil interacting with vortex gusts \citep{fukami2024data}.

Despite the advantages of the phase-reduction technique in studying and controlling oscillatory flows, most previous studies have focused on two-dimensional laminar flows with perfect periodicity. The phase-reduction analysis is theoretically founded on periodic systems with a limit cycle solution; therefore, the conventional phase-reduction framework is limited in its application to chaotic oscillators. However, most oscillatory flows in practical engineering problems are not perfectly periodic and exhibit chaotic behavior with multiple frequencies and broadband spectra, stemming from their complex nonlinear dynamics accompanied by flow instabilities and turbulence. To overcome this limitation, several studies have extended phase-reduction analysis to chaotic flows with a broadband frequency spectrum based on the ensemble-averaging approach \citep{kim2024influence} and the pulse-train method \citep{godavarthi2025phase}. These studies have demonstrated that the phase-reduction analysis supplemented with appropriate statistical treatments is capable of addressing a wider range of unsteady oscillatory flows, which are not necessarily perfectly periodic. Nonetheless, there still exist opportunities to establish phase-reduction analysis for turbulent flows with multiple dominant frequencies. Furthermore, characterizing the phase variable of turbulent flows with multiple dominant frequencies is even challenging within existing phase-reduction methods.

In this study, we develop a data-driven framework for oscillator-based modeling that learns phase-amplitude coordinates of turbulent flows from flow field data and their temporal dynamics. To address the challenges in the analytic application of the phase-reduction approach to multi-frequency flows, machine-learning techniques are adopted to represent the multi-frequency dynamics of the turbulent flow with a set of oscillators. Moreover, the dynamics of the extracted oscillators are modeled using a data-driven method to offer an accurate prediction of the oscillatory behavior in turbulent flows.

This paper is organized as follows. In Sec.~\ref {sec:2}, we begin our discussion with a brief introduction to the oscillator-based description of unsteady fluid flows. Sec.~\ref{sec:2} also provides detailed explanations of the proposed data-driven framework based on autoencoders and the neural ODE for oscillator-based modeling of turbulent flow dynamics with multiple dominant frequencies. We demonstrate the proposed approach for a supersonic turbulent cavity flow in Sec.~\ref{sec:3}. Conclusions are provided in Sec.~\ref{sec:4}.

\section{Oscillator-based modeling of turbulent flows}    \label{sec:2}

\subsection{Oscillator-based description of turbulent flows with multiple dominant frequencies}    \label{sec:2-1}

Let us propose a data-driven approach for reduced-order modeling of turbulent flows with multiple dominant frequencies using oscillators. The dynamics of fluid flows are governed by the Navier--Stokes (NS) equations, which can be compactly expressed as
\begin{equation}    \label{eq:GoverningEq}
    \frac{\partial}{\partial t} \bs{q}(\bs{x},t) = \bs{\mathcal{N}}(\bs{q}(\bs{x},t)),
\end{equation}
where $\bs{q}$ is the flow state vector, $\bs{\mathcal{N}}$ denotes the Navier--Stokes operator, and $\bs{x}$ and $t$ represent the spatial and temporal variables, respectively.

When the flow is perfectly periodic with the natural angular frequency $\Omega$, we can consider it as a limit cycle solution $\bs{q}_0$ of the governing equation Eq.~\ref{eq:GoverningEq}, which satisfies $\bs{q}_0(\bs{x},t) = \bs{q}_0(\bs{x},t+2\pi/\Omega)$. To simplify the description of the flow, we introduce the phase functional $\Theta[\bs{q}]$ that maps the flow field $\bs{q}$ to the phase variable $\theta = \Theta[\bs{q}]$. The phase functional can be further extended to the vicinity of the limit cycle solution to represent the weakly perturbed flow. The phase dynamics of the unperturbed periodic flow can be derived from its governing equation Eq.~\ref{eq:GoverningEq},
\begin{equation}    \label{eq:PhaseDynamics_PeriodicFlow}
    \dot{\theta} = \dot{\Theta}[\bs{q}]
                 = \int_{\mathcal{V}_f} \frac{\delta\Theta}{\delta\bs{q}} \cdot \dot{\bs{q}} \, \mathrm{d}\bs{x}
                 = \int_{\mathcal{V}_f} \frac{\delta\Theta}{\delta\bs{q}} \cdot \bs{\mathcal{N}}(\bs{q}) \, \mathrm{d}\bs{x}
                 = \Omega,
\end{equation}
where $\delta\Theta/\delta\bs{q}$ denotes the functional derivative of $\Theta(\bs{q})$, and $\mathcal{V}_f$ represents the control volume of interest. When the flow is weakly perturbed, this formalism enables characterization of the asymptotic influence of external perturbations on the phase dynamics of the flow by the phase-sensitivity function. As a result, the phase-based description of the fluid flow reveals the phase response to external forcing, providing a deeper insight into the perturbation dynamics of periodic flows \citep{nakao2016phase, taira2018phase}.

The aforementioned phase-reduction formulation is only valid for perfectly periodic flows. Therefore, it is not straightforward to apply the phase-reduction to oscillatory flows with chaotic characteristics due to their broadband frequency spectrum. However, when a single frequency dominates the flow physics, we can still employ the phase-based description approximately by incorporating stochasticity in the phase dynamics as
\begin{equation}    \label{eq:PhaseDynamics_OscillatoryFlow_SingleFreq}
    \dot{\theta} = \int_{\mathcal{V}_f} \frac{\delta\Theta}{\delta\bs{q}} \cdot \bs{\mathcal{N}}(\bs{q}) \, \mathrm{d}\bs{x}
                 = \Omega + n_{\theta}(t),
\end{equation}
where $n_\theta(t)$ represents the instantaneous fluctuation in the temporal variation of the phase variable associated with the flow dynamics unresolved by the phase-based description.

Alongside the phase dynamics of the flow, the amplitude variable $r$ can also be considered to supplement the description of the flow dynamics. The amplitude variable of unsteady flows can be defined in various ways depending on its objectives. \cite{fukami2024data} characterized the deviation of the perturbed flow from the stable periodic solution by the amplitude variable defined based on the Koopman theory, which is a standard method for periodic dynamical systems with a limit cycle \citep{mauroy2020koopman}. \cite{kim2024influence} and \cite{godavarthi2025phase} embedded a physical quantity associated with the coherence of vortical structures into the amplitude variable to reflect the instantaneous chaotic characteristics of the flow. In this study, we define the amplitude variable to capture the strength of the oscillatory behavior in the flow. The flow field $\bs{q}(\bs{x}, t)$ can be linearly decomposed into the temporal mean $\bar{\bs{q}}(\bs{x})$ and the fluctuating component $\bs{q}^{\prime}(\bs{x},t)$. We now define the amplitude variable of the flow as
\begin{equation}    \label{eq:Amplitude_PeriodicFlow}
    r = \left( \int_{\mathcal{V}_f} \bs{q}^\prime(\bs{x}, t) \cdot \bs{q}^\prime(\bs{x}, t) \mathrm{d}\bs{x} \right)^{1/2},
\end{equation}
which quantifies the energy of the fluctuating component $\bs{q}^{\prime}$ of the flow with the assumption that a single amplitude variable fully captures the fluctuating flow components.

Analogous to Eq.~\ref{eq:PhaseDynamics_PeriodicFlow}, the amplitude dynamics of unperturbed periodic flows can be written as
\begin{equation}    \label{eq:AmplitudeDynamics_PeriodicFlow}
    \dot{r} = \dot{R}[\bs{q}]
            = \int_{\mathcal{V}_f} \frac{\delta R}{\delta\bs{q}} \cdot \dot{\bs{q}} \, \mathrm{d}\bs{x}
            = \int_{\mathcal{V}_f} \frac{\delta R}{\delta\bs{q}} \cdot \bs{\mathcal{N}}(\bs{q}) \, \mathrm{d}\bs{x}
            = \Psi(\theta, r),
\end{equation}
where $R[\bs{q}]$ denotes the amplitude functional that returns the amplitude variables from the flow field. We note that the amplitude variable in the current formulation reflects the unsteady characteristics of the flow and differs from the amplitude definition based on Koopman theory, which measures the deviation of the system from its limit cycle \citep{mauroy2020koopman, namura2022estimating}. As a result, the oscillators representing periodic flows behave similarly to the Stuart-Landau oscillator. When the periodic flow is perturbed, the amplitude variable initially deviates and gradually approaches the reference amplitude of the limit cycle solution as the flow state $\bs{q}$ converges to the limit cycle solution and the transient behavior vanishes. Eq.~\ref{eq:AmplitudeDynamics_PeriodicFlow} can be approximately extended to chaotic oscillatory flows with a single dominant frequency as
\begin{equation}    \label{eq:AmplitudeDynamics_OscillatoryFlow_SingleFreq}
    \dot{r} = \int_{\mathcal{V}_f} \frac{\delta R}{\delta\bs{q}} \cdot \bs{\mathcal{N}}(\bs{q}) \, \mathrm{d}\bs{x}
            = \Psi(\theta, r) + n_{r}(t),
\end{equation}
where $n_r(t)$ denotes the temporal fluctuation in the amplitude dynamics stemming from the unresolved flow dynamics.

To address multiple dominant frequencies in turbulent flows, the phase-amplitude-reduction formulation can be further extended by introducing a set of coupled oscillators. If the turbulent flow exhibits $M$ distinct dominant frequencies with isolated bands in the spectrum, the fluctuation of the flow $\bs{q}^{\prime}(\bs{x},t)$ around its temporal average $\bar{\bs{q}}(\bs{x})$ can be approximately decomposed into modal contributions $\bs{q}^{\prime}_m$,
\begin{equation}    \label{eq:FlowDecomposition}
    \bs{q}(\bs{x},t) = \bar{\bs{q}}(\bs{x}) + \bs{q}^{\prime}(\bs{x},t) \approx \bar{\bs{q}}(\bs{x}) + \sum_{m=1}^M \bs{q}^{\prime}_m(\bs{x},t),
\end{equation}
where $m$ is the frequency index. We then consider $M$ oscillators corresponding to the flow modes, such that their phase and amplitude variables can be described as
\begin{equation}    \label{eq:PhaseDynamics_OscillatoryFlow_MultiFreq}
    \dot{\bs{\theta}} = \int_{\mathcal{V}_f} \frac{\delta\bs{\Theta}}{\delta\bs{q}} \cdot \bs{\mathcal{N}}(\bs{q}) \, \mathrm{d}\bs{x}
                 = \bs{\Omega} + \bs{g_{\theta}}(\bs{\theta},\bs{r}) + \bs{n_{\theta}}(t),
\end{equation}
\begin{equation}    \label{eq:AmplitudeDynamics_OscillatoryFlow_MultiFreq}
    \dot{\bs{r}} = \int_{\mathcal{V}_f} \frac{\delta \bs{R}}{\delta\bs{q}} \cdot \bs{\mathcal{N}}(\bs{q}) \, \mathrm{d}\bs{x}
                 = \bs{\Psi}(\bs{\theta},\bs{r}) + \bs{g_r}(\bs{\theta},\bs{r}) + \bs{n_r}(t),
\end{equation}
which is the extension of oscillator-based description to the flows with $M$ dominant frequencies, Eqs.~\ref{eq:PhaseDynamics_OscillatoryFlow_MultiFreq} and~\ref{eq:AmplitudeDynamics_OscillatoryFlow_MultiFreq}. The phase and amplitude vector stores the phase and amplitude of $M$ oscillators as

\begin{equation}
\bs{\theta}(\bs{q}(\bs{x},t)) \equiv \begin{bmatrix} \theta_1(\bs{q}(\bs{x},t)) \\
                                                     \theta_2(\bs{q}(\bs{x},t)) \\
                                                     \vdots                     \\
                                                     \theta_M(\bs{q}(\bs{x},t))
                                     \end{bmatrix}, \quad
\bs{r}(\bs{q}(\bs{x},t)) \equiv \begin{bmatrix} r_1(\bs{q}(\bs{x},t)) \\
                                                r_2(\bs{q}(\bs{x},t)) \\
                                                \vdots                \\
                                                r_M(\bs{q}(\bs{x},t))
                                \end{bmatrix}.
\end{equation}
The amplitude variable $r_m$ of each oscillator is defined by extending the definition in Eq.~\ref{eq:Amplitude_PeriodicFlow} as
\begin{equation}    \label{eq:Amplitude_OscillatoryFlow_MultiFreq}
    r_m(t) = \left( \int_{\mathcal{V}_f} \bs{q}_{m}^{\prime}(\bs{x},t) \cdot \bs{q}_{m}^{\prime}(\bs{x},t) \mathrm{d}\bs{x} \right)^{1/2},
\end{equation}
which quantifies the strength of each flow mode. The coupling functions $\bs{g_\theta}$ and $\bs{g_r}$ account for interactions among oscillators due to the nonlinearity of flow physics, while $\bs{\Omega}$ and $\bs{\Psi}$ represent the intrinsic oscillator frequencies and amplitude dynamics, respectively.

\subsection{Oscillator extraction by oscillator-identifying autoencoders}    \label{sec:2-2}

\begin{figure*}
    \centering
    \includegraphics[width=1.0\textwidth]{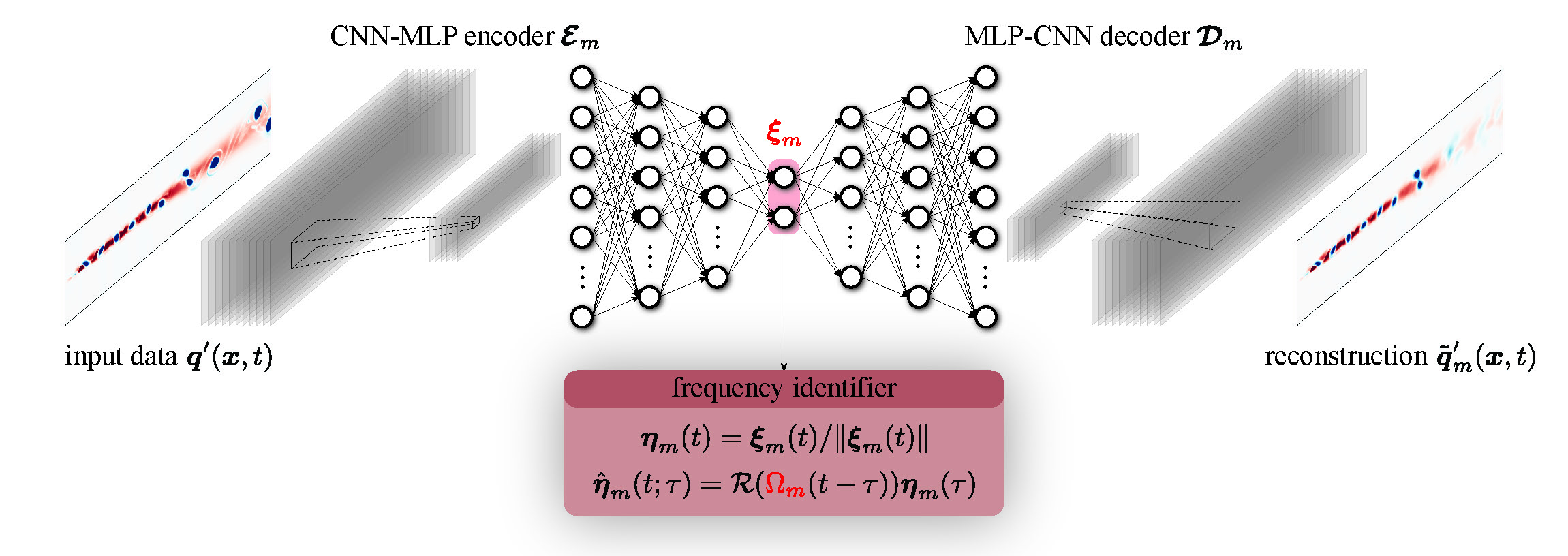}
    \caption{Architecture of a single oscillator-identifying autoencoder with a frequency identifier.
    \label{fig01:SingleAE}}
\end{figure*}

Analytic derivations of the phase and amplitude functionals for multi-frequency fluid flows are generally not accessible due to the nonlinearity in the governing equation Eq.~\ref{eq:GoverningEq}. Therefore, alternative approaches for characterizing oscillators are necessary for the practical applications of phase-reduction to multi-frequency fluid flows. Most implementations of phase-reduction for studying single-frequency flows rely on observables to define the oscillator, such as local probe measurements \citep{loe2023controlling} and forces on a bluff body~\citep{khodkar2020phase, kim2024influence}. Alternatively, the entire flow field can be used to extract the oscillator. For instance, oscillators can be characterized by orthogonal projections of the flow field onto DMD modes~\citep{godavarthi2025phase}. Nonlinear autoencoders can discover a low-order manifold in which oscillatory flows appear as cyclic trajectories by compressing flow field data to the latent space, enabling the definition of the oscillator~\citep{fukami2024data}.

\begin{figure*}
    \centering
    \includegraphics[width=1.0\textwidth]{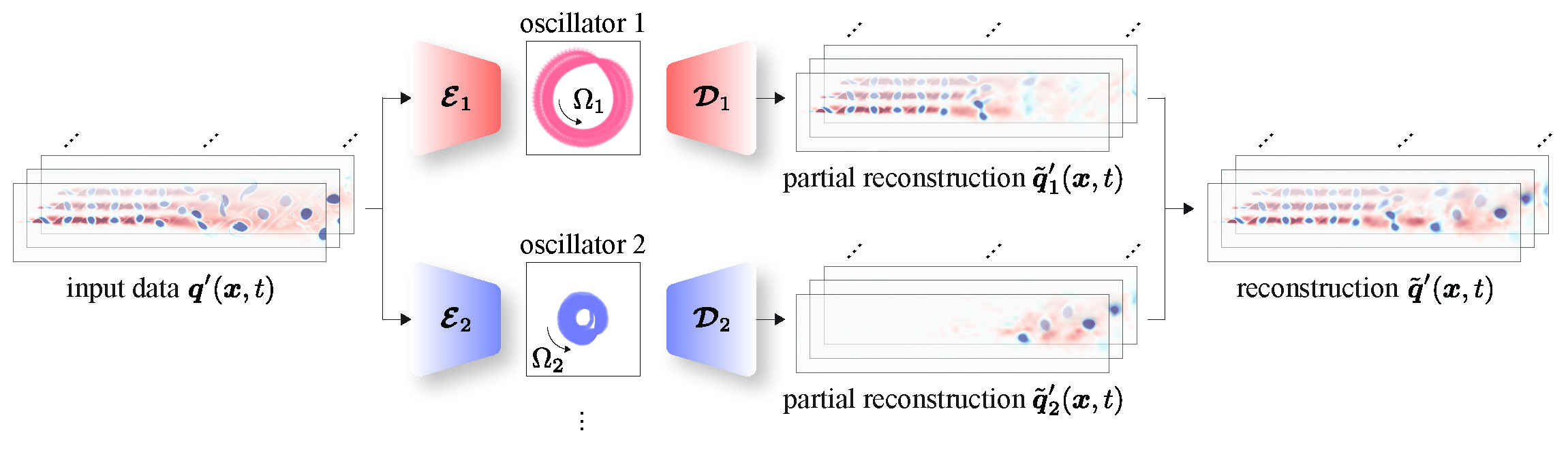}
    \caption{Parallel arrangement of multiple oscillator-identifying autoencoders extracting two oscillators.
    \label{fig02:MultipleAE}}
\end{figure*}

In recent studies, the phase-autoencoder has been developed as a data-driven approach for phase-reduction of various dynamical systems \citep{yawata2024phase,yawata2025phase}. The phase-autoencoder directly approximates the phase functional of periodic dynamical systems by structuring its latent space as a two-dimensional oscillator. It has been demonstrated that the phase-autoencoder can successfully extract oscillators from various dynamical systems, ranging from simple low-dimensional oscillators to complex spatiotemporal patterns. Inspired by these efforts, we propose a data-driven approach to extract a set of representative oscillators from the time series of a turbulent flow with multiple dominant frequencies based on oscillator-identifying autoencoders, as provided in Figs.~\ref{fig01:SingleAE} and \ref{fig02:MultipleAE} illustrated with results of a simple example with a two-dimensional laminar free shear layer \citep{yeh2017laminar}. A single oscillator-identifying autoencoder is composed of an encoder $\bs{\mathcal{E}}_m$ and a decoder $\bs{\mathcal{D}}_m$, constructed by convolutional neural networks (CNN) and multilayer perceptrons (MLP). To extract oscillators associated with different dominant frequencies separately, multiple oscillator-identifying autoencoders are arranged in parallel. Similar parallel arrangements have been adopted in previous studies on nonlinear modal representations of fluid flows \citep{murata2020nonlinear, fukami2020convolutional}. The encoder compresses the fluctuating component of the flow field data $\bs{q}^{\prime} = \bs{q} - \bar{\bs{q}}$ to a two-dimensional latent oscillator $\bs{\xi}_m = [\xi^{(1)}_m\ \xi^{(2)}_m]^{\intercal} = \bs{\mathcal{E}}_m(\bs{q}^{\prime}(\bs{x},t))$. Here, the flow field data $\bs{q}$ is taken to be the windowed state of the flow, which retains the essential oscillatory behavior of the flow and lowers the computational cost to train autoencoders. The encoder characterizes the phase and amplitude variables of the flow as
\begin{equation}\label{eq:PhaseCharacterization}
    \theta_m = \Theta_m(\bs{q}(\bs{x},t)) = \angle\bs{\xi}_m(\bs{q}(\bs{x},t)) = \arg(\xi^{(1)}_m + i\xi^{(2)}_m),
\end{equation}
\begin{equation}\label{eq:AmplitudeCharacterization}
    r_m = R_m(\bs{q}(\bs{x},t)) = \| \bs{\xi}_m(\bs{q}(\bs{x},t)) \| = [(\xi^{(1)}_m)^2 + (\xi^{(2)}_m)^2]^{1/2}.
\end{equation}
Since oscillators inherit distinctive frequencies from the flow, the decoder partially reconstructs the flow field by extracting the modal contribution as $\tilde{\bs{q}}^{\prime}_{m}(\bs{x},t) = \bs{\mathcal{D}}_m(\bs{\xi}_m) = \bs{\mathcal{D}}_m(\bs{\mathcal{E}}_m(\bs{q}^{\prime}(\bs{x},t)))$. By superposing the outputs $\tilde{\bs{q}}^{\prime}_{m}$ of all oscillator-identifying autoencoders, the final reconstruction of the flow field is calculated as
\begin{equation}    \label{eq:FinalReconstruction}
    \tilde{\bs{q}}(\bs{x},t) \approx \bar{\bs{q}}(\bs{x}) + \tilde{\bs{q}}^{\prime}(\bs{x},t)
    = \bar{\bs{q}}(\bs{x}) + \sum_{m = 1}^M\tilde{\bs{q}}^{\prime}_m(\bs{x},t)
    = \bar{\bs{q}}(\bs{x}) + \sum_{m=1}^M\bs{\mathcal{D}}_m(\bs{\mathcal{E}}_m(\bs{q}^{\prime}(\bs{x},t))),
\end{equation}
which follows Eq.~\ref{eq:FlowDecomposition}.

Training a group of oscillator-identifying autoencoders requires careful guidance due to the large number of trainable parameters and the non-convexity of the optimization problem. To stabilize the training procedure, we sequentially introduce and train the oscillator-identifying autoencoders. After training the $(m - 1)$-th phase autoencoder, we add the $m$-th oscillator-identifying autoencoder to the arrangement and optimize it while keeping all previously trained parameters fixed. This training procedure is iterated until $M$ representative oscillators are obtained corresponding to the desired dominant frequencies.

For the optimization of an oscillator-identifying autoencoder, three training stages are designed to form the latent space into an oscillator with desired behavior and properties. We begin with the pretraining of the oscillator-identifying autoencoder using the loss function $\mathcal{L}_{\bs{q}_m}$ for the flow field reconstruction only,
\begin{equation}    \label{eq13:ReconstructionLoss_AE}
    \mathcal{L}_{m,1} = \mathcal{L}_{\bs{q}_m} = \sum_{l=1}^{N_s} \| \bs{q}^{\prime}(\bs{x},t_l) - \sum_{j = 1}^{m} \tilde{\bs{q}}^{\prime}_j(\bs{x},t_l) \|^2,
\end{equation}
where $\|\cdot\|$ represents the L2 norm and $N_s$ is the number of snapshots in the randomly sampled training batch. Through pretraining, the oscillator-identifying autoencoder first converges to the structure that optimally captures the flow features and establishes a preliminary latent space. As a result, the pretrained autoencoder suggests the target frequency at each stage from its partial reconstruction of the data and the frequency analysis of latent variables, determining the extraction order of representative oscillators, which is closely connected to the overall dominance of the frequencies.

The following stage fine-tunes the oscillator-identifying autoencoder to learn the phase variable by shaping the latent space as an oscillator with a globally consistent angular speed by imposing a phase loss term. To formulate the phase loss, we introduce a frequency identifier with a trainable parameter $\Omega_m$, attached to the latent space of the oscillator-identifying autoencoder to capture the mean angular speed of the oscillator during training. Based on $\Omega_m$, the frequency identifier calculates an ideal evolution of the normalized latent vector $\bs{\eta}_m(t) = \bs{\xi}_m(t)/\|\bs{\xi}_m(t)\|$ from the reference time $t = \tau$ by applying a rotation matrix $\mathcal{R}$ as follows;
\begin{equation}
    \hat{\bs{\eta}}_m(t;\tau) = \mathcal{R}(\Omega_m(t - \tau))\bs{\eta}_m(\tau),    \quad    \mathcal{R}(\phi) = \begin{bmatrix} \cos\phi & -\sin\phi \\ \sin\phi & \cos\phi \end{bmatrix}.
\end{equation}

In order to train the frequency identifier and update the oscillator-identifying autoencoder, we employ two phase loss formulations. We first consider the global form of the phase loss $\mathcal{L}_{\theta_m}^\mathrm{G}$ defined as
\begin{equation}    \label{eq:PhaseLoss_Strong_AE}
    \mathcal{L}_{\theta_m}^{\mathrm{G}} = \beta_{\theta_m}^\mathrm{G}\sum_{l=1}^{N_s} \| \bs{\eta}_{m}(t_l) - \hat{\bs{\eta}}_{m}(t_l;t_1) \|^2,
\end{equation}
leading to the loss function $\mathcal{L}_{m,2}^\mathrm{G} = \mathcal{L}_{\bs{q}_m} + \mathcal{L}_{\theta_m}^\mathrm{G}$ combined with the reconstruction loss, where $\beta_{\theta_m}^\mathrm{G}$ is a constant weighting hyperparameter to balance the loss terms. The global phase loss $\mathcal{L}_{\theta_m}^\mathrm{G}$ compares the normalized latent vectors directly obtained from the oscillator-identifying autoencoder to the ideal evolution from the reference point. For training with the global phase loss, the training batches are constructed by sampling contiguous subsequences from the full training data. This enables the frequency identifier to capture the mean angular frequency coherent over the local data segment and strongly promotes the latent vector to behave as an oscillator. The reference time $\tau$ to evaluate the ideal evolution $\hat{\bs{\eta}}_m$ is chosen as the time of the first snapshot in the training batch $\tau = t_1$.

Since the global phase loss assumes a linear phase evolution of the oscillator over a segment from the reference point, it can suppress the local fluctuation of the angular velocity in the latent space. Therefore, after training with the global phase loss, we convert the phase loss to a local form defined as
\begin{equation}    \label{eq:PhaseLoss_Weak_AE}
    \mathcal{L}_{\theta_m}^\mathrm{L} = \sum_{l=1}^{N_s/2} \beta_{\theta_m}^\mathrm{L}\| \bs{\eta}_{m}(t_{2l}) - \hat{\bs{\eta}}_{m}(t_{2l};t_{2l-1})\|^2.
\end{equation}
which yields the loss function $\mathcal{L}_{m,2}^\mathrm{L} = \mathcal{L}_{\bs{q}_m} + \mathcal{L}_{\theta_m}^\mathrm{L}$ with the balancing parameter $\beta_{\theta_m}^\mathrm{L}$. Here, the local phase loss uses pairwise batches that randomly sample pairs of consecutive snapshots $\bs{q}(\bs{x},t_l)$ and $\bs{q}(\bs{x},t_{l+1}) = \bs{q}(\bs{x},t_l+\Delta t)$ from the training data, where $\Delta t$ is the time step size of the time series data. The conversion to the weak phase loss provides flexibility to the variation of the local angular speed in the latent space, promoting smoothness in the oscillator behavior.

However, training an oscillator-identifying autoencoder with a phase loss term can degrade the generalization performance when the strength of the oscillatory component becomes weak. When the energy of the oscillatory mode approaches zero, the phase variable of the oscillator can be ill-defined, causing a large phase loss. To reduce the phase loss, the optimizer tends to overly extract the flow features from the data and sacrifice the reconstruction accuracy during training. To resolve this issue, we employ a mask function to relax the regularization by the phase loss by imposing small weights on the phase loss when the oscillatory component is not sufficiently strong to define the phase variable. We use a transformed sigmoid function for the weight parameter $\beta_{\theta_m}^\mathrm{L}$, defined as
\begin{equation}    \label{eq:MaskFunction}
    \beta_{\theta_m}^\mathrm{L} = \frac{\gamma_{\theta_m}}{1 + e^{-a_m(E_m-b_m)}},    \quad    E_m(\tilde{\bs{q}}_{m}^{\prime}(\bs{x},t)) = \left(\int_{\mathcal{V}_f} \tilde{\bs{q}}_{m}^{\prime}(\bs{x},t) \cdot \tilde{\bs{q}}_{m}^{\prime}(\bs{x},t) \mathrm{d}\bs{x}\right)^{1/2}.
\end{equation}
The hyperparameter $\gamma_{\theta_m}$ rescales the sigmoid function, and $a_m$ and $b_m$ control the slope and translation of the mask function, respectively.

As the final step, we introduce the constraint on the amplitude variable to capture the strength of the modal contribution to the flow field, as defined in Eq.~\ref{eq:Amplitude_OscillatoryFlow_MultiFreq}. We again fine-tune the oscillator-identifying autoencoder by adding the amplitude term $\mathcal{L}_{r_m}$ to the loss function, defined as
\begin{equation}    \label{eq:AmplitudeLoss_AE}
    \mathcal{L}_{r_m} = \sum_{l=1}^{N_s} \beta_{r_m}\left[(r_m(t_l) - E_m(\tilde{\bs{q}}_{m}^{\prime}(\bs{x},t_l))\right]^2
\end{equation}
with the weighting hyperparameter $\beta_{r_m}$ for the amplitude loss. The final loss function to optimize the $m$-th oscillator-identifying autoencoder is $\mathcal{L}_{m,3} = \mathcal{L}_{\bs{q}_m} + \mathcal{L}_{\theta_m}^\mathrm{L} + \mathcal{L}_{r_m}$.

As a consequence of the following fine-tuning stages by introducing phase and amplitude loss terms, the reconstruction loss increases compared to that of the pretraining stage. This increased reconstruction loss is attributed to the autoencoder dropping irrelevant flow features as it shapes its latent state as an oscillator. Therefore, the fine-tuning stages can be considered as a conditioning of the autoencoder to concentrate on learning flow features associated with the target frequency and the mapping between the flow field data and the oscillator by implementing regularizations.

We note that the current oscillator-based description of turbulent flows utilizing autoencoders is distinctive from the conventional phase-amplitude-reduction analysis. As mentioned earlier in Sec.~\ref{sec:2-2}, the phase-amplitude-reduction analysis is established for systems with a stable limit cycle solution. Since turbulent flows do not exhibit limit cycle solutions, direct application of the conventional framework with the introduction of phase-reduction objects forces challenges. On the other hand, the oscillator-identifying autoencoders find representative oscillators of the turbulent flow with multiple dominant frequencies by regularizing the latent space with phase dynamics and amplitude, providing an alternative viewpoint for turbulent flows based on oscillators inspired by the conventional phase-amplitude-reduction method.

\subsection{Latent oscillator dynamics modeling and data assimilation}    \label{sec:2-3}

Modeling dynamics of representative oscillators enables us to predict the oscillatory behavior of the turbulent flow associated with its dominant frequencies. Recently, data-driven approaches equipped with neural networks have been widely employed in modeling dynamical systems. In particular, the neural ordinary differential equation (neural ODE) developed by \cite{chen2018neural}, which combines a neural network and a numerical solver for ordinary differential equations (ODEs), has gained attention owing to its high capacity to learn the nonlinear dynamics of systems \citep{linot2023dynamics, linot2023stabilized, chakraborty2024divide}. In contrast to the family of recurrent neural networks, the neural ODE represents the dynamical system in continuous time, allowing adaptive time resolution in model evaluation. Motivated by these advantages, we utilize neural ODE to learn the dynamics of oscillators in turbulent flows extracted by oscillator-identifying autoencoders.

\begin{figure*}
    \centering
    \includegraphics[width=1.0\textwidth]{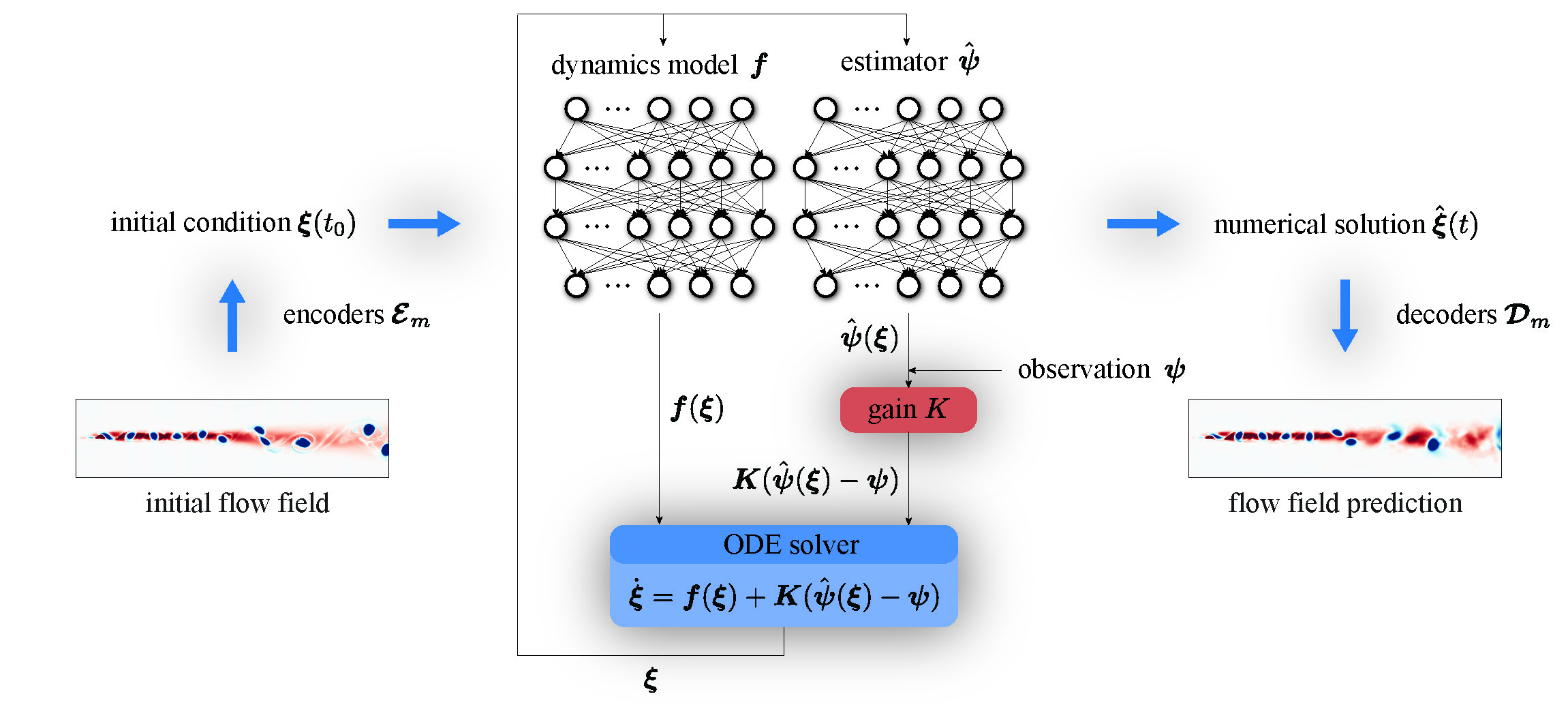}
    \caption{A schematic of a neural ODE to learn and predict oscillator dynamics.
    \label{fig03:NeuralODE}}
\end{figure*}

The dynamics of coupled oscillators in turbulent flows with multiple dominant frequencies given in Eqs.~\ref{eq:PhaseDynamics_OscillatoryFlow_MultiFreq} and \ref{eq:AmplitudeDynamics_OscillatoryFlow_MultiFreq} can be compactly rewritten in terms of $\bs{\xi}$ through the coordinate transformation,
\begin{equation}    \label{eq:LatentDynamics}
    \dot{\bs{\xi}} = \bs{f} \left(\bs{\xi}\right),
\end{equation}
where $\bs{\xi} = [\bs{\xi}_1^{\intercal} \ \bs{\xi}_2^{\intercal} \ \cdots \ \bs{\xi}_M^{\intercal}]$ collects all oscillator variables obtained from the oscillator-identifying autoencoders. To model the right-hand side of Eq.~\ref{eq:LatentDynamics}, we employ an MLP as illustrated in Fig.~\ref{fig03:NeuralODE}, which offers a prediction $\hat{\bs{\xi}}$ through numerical integration in time by an ODE solver for a given initial condition. The dynamics function $\bs{f}(\bs{\xi})$ can be trained directly by regression of the temporal derivative $\dot{\bs{\xi}}$ using the loss function
\begin{equation}    \label{eq:LossFunction_NODE_Regression}
    \mathcal{L}_{\bs{f}} = \sum_{l=1}^{N_t}\| \bs{f}(\bs{\xi}(t_l)) - \dot{\bs{\xi}}(t_l)\|^2,
\end{equation}
where $\dot{\bs{\xi}}$ is evaluated by numerical differentiation. Alternatively, numerical solutions of Eq.~\ref{eq:LatentDynamics} can be used to learn the function $\bs{f}(\bs{\xi})$. In this case, the loss function compares the numerical solutions with the original oscillator variables for $N_r$ time steps, which can be written as
\begin{equation}    \label{eq:LossFunction_NODE_Rollout}
    \mathcal{L}_{\bs{f}} = \sum_{l=1}^{N_s}\sum_{n=1}^{N_r}\| \bs{\xi}(t_l + n\Delta t) - \hat{\bs{\xi}}(t_l + n\Delta t;\bs{\xi}(t_l),t_l)\|^2,
\end{equation}
where $\hat{\bs{\xi}}(t_0 + T;\bs{\xi}_0,t_0)$ is the numerical solution at $t = t_0 +T$ for the initial condition $\bs{\xi}=\bs{\xi}_0$ at $t = t_0$.

While the dynamics function $\bs{f}(\bs{\xi})$ is trained successfully and accurately predicts the oscillator behavior for short-terms, long-term predictions of the oscillator dynamics eventually deviate from the ground-truth solution due to chaotic characteristics and unresolved dynamics of the flow, especially for turbulent flows. To improve the accuracy of the long-term prediction, a correction term is added to Eq.~\ref{eq:LatentDynamics} based on the observation $\bs{\psi}$ and its estimation, leading to
\begin{equation}    \label{eq:LatentDynamics_wCorrection}
    \dot{\bs{\xi}} = \bs{f} \left(\bs{\xi}\right) + \bs{K}[\hat{\bs{\psi}}(\bs{\xi}) - \bs{\psi}],
\end{equation}
where $\hat{\bs{\psi}}$ and $\bs{K}$ denote the observable estimator and the gain function. We note that the observation $\bs{\psi}$ is generally not assumed to be part of the flow field data $\bs{q}$ used for training autoencoders, and it can be any other flow measurements at appropriate locations. To model the correction term, we first train the estimator $\hat{\bs{\psi}}$ by regression using the loss function
\begin{equation}    \label{eq:LossFunction_Estimator}
    \mathcal{L}_{\hat{\bs{\psi}}} = \sum_{l=1}^{L}\| \hat{\bs{\psi}}(\bs{\xi}(t_l)) - \bs{\psi}(t_l)\|^2.
\end{equation}
Later, gain function $\bs{K}$ is learned with fixed $\bs{f}$ and $\hat{\bs{\psi}}$ using the same loss function Eq.~\ref{eq:LossFunction_NODE_Rollout}, where the numerical solution $\hat{\bs{\xi}}$ is obtained by solving Eq.~\ref{eq:LatentDynamics_wCorrection}. In place of directly learning $\bs{K}$, an ensemble Kalman filter (EnKF) can be utilized as considered by \cite{mousavi2025sequential}. Although EnKF provides optimal estimates of the latent state with uncertainty quantification, it may require a large number of ensembles and a priori known statistics of the model and noise for the propagation and innovation processes. In contrast, the pretrained gain function $K$ in this study can provide direct corrections to the prediction of latent oscillators without ensemble-based propagation and statistical information. Once the neural networks $\bs{f}$, $\hat{\bs{\psi}}$ and $\bs{K}$ are trained, oscillator behavior prediction only requires external observations $\bs{\psi}$ to solve the learned ordinary differential equation, not the full flow field used to train autoencoders. In addition, numerical solution $\hat{\bs{\xi}}$ can provide the flow field reconstruction $\hat{\bs{q}}$ using the trained decoders as
\begin{equation}
    \hat{\bs{q}}(\bs{x},t)
    = \bar{\bs{q}}(\bs{x}) + \hat{\bs{q}}'(\bs{x},t)
    = \bar{\bs{q}}(\bs{x}) + \sum_{m=1}^{M}\hat{\bs{q}}_m'(\bs{x},t)
    = \bar{\bs{q}}(\bs{x}) + \sum_{m=1}^{M}\bs{\mathcal{D}}_m(\hat{\bs{\xi}}_m(t)),
\end{equation}
where $\hat{\bs{q}}_m'$ is the partial reconstruction of the flow field based on the numerical solution of the $m$-th oscillator.

\section{Applications}    \label{sec:3}

\begin{figure*}
    \centering
    \includegraphics[width=1.0\textwidth]{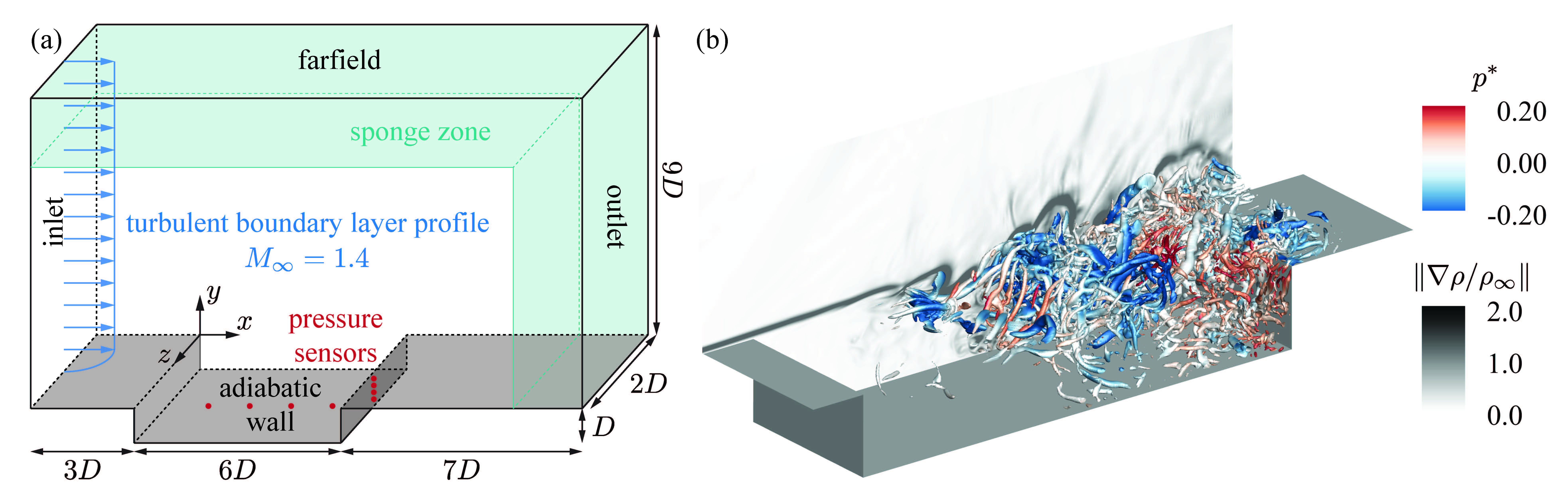}
    \caption{(a) Computational domain for LES of the three-dimensional supersonic turbulent cavity flow. Red circles represent the location of pressure sensor measurements used for the observation-based correction term in the dynamics modeling. (b) Instantaneous vortical structures visualized by $Q$-criterion ($QD^2/u^2_{\infty} = 15$) colored by the nondimensionalized pressure $p^*$ with the magnitude of density gradient $\nabla\rho/\rho_{\infty}$ on the plane $z/D = -1$.
    \label{fig04:ComputationalDomain_3DCF}}
\end{figure*}

To demonstrate the data-driven framework proposed in Sec.~\ref{sec:2}, we consider a three-dimensional supersonic turbulent flow over a cavity \citep{liu2021unsteady, godavarthi2025phase} illustrated in Fig.~\ref{fig04:ComputationalDomain_3DCF}(a). This canonical turbulent flow contains multiple dominant frequencies. This example considers a rectangular cavity with an aspect ratio of $L/D = 6$, where $L$ and $D$ are the length and depth of the cavity, respectively. The freestream has the Mach number of $M_{\infty} = u_{\infty}/a_{\infty} = 1.4$ and the Reynolds number of $\mathrm{Re}_D = \rho_{\infty}u_{\infty}D/\mu_\infty=10^4$ based on the depth of the cavity, where $a_{\infty}$, $u_\infty$, $\rho_\infty$ and $\mu_\infty$ represent the speed of sound, velocity, density and dynamic viscosity of the freestream, respectively. The Prandtl number and the specific heat ratio are chosen as $\mathrm{Pr} = \rho_\infty\mu_\infty/\alpha_\infty = 0.7$ and $\gamma = 1.4$, respectively, based on the standard values of the air, where $\alpha_{\infty}$ is the thermal diffusivity of the freestream.

To gain access to the flow field, we perform a large-eddy simulation (LES) using a compressible flow solver \textit{CharLES} \citep{khalighi2011noise, khalighi2011unstructured}, which adopts a second-order finite-volume method for spatial discretization and a third-order Runge-Kutta scheme for time integration. We utilize the Vreman model \citep{vreman2004eddy} for the sub-grid scale turbulence and the Harten-van-Leer scheme \citep{toro1994restoration} to capture shocks. We use a Cartesian grid for the computational domain, $(x,y,z)/D \in [-3,13] \times [-1,8] \times [-1,1]$ with the origin placed at the leading edge of the cavity, as shown in Fig.~\ref{fig04:ComputationalDomain_3DCF}(a). The dynamic viscosity is calculated by a power law $\mu(T) = \mu_{\infty}(T/T_{\infty})^{0.76}$ with a reference temperature $T_{\infty}$ of the freestream, which approximates Sutherland's law \citep{garnier2009large}.

The velocity at the inlet boundary is prescribed by the 1/7th power law of the turbulent boundary layer superposed with the turbulent fluctuation modeled by random Fourier modes \citep{bechara1994stochastic}, resulting in an initial boundary layer thickness of $\delta_0/D = 0.167$ at the leading edge. An adiabatic no-slip condition is assigned to the cavity walls, and the periodic boundary condition is enforced in the spanwise direction. A sponge zone is applied to the outlet and farfield boundaries to damp out the pressure waves and avoid their numerical reflections.

An instantaneous vortical field obtained from the flow simulation is visualized in Fig.~\ref{fig04:ComputationalDomain_3DCF}(b) using an isosurface of the $Q$-criterion, colored by the nondimensionalized pressure defined as $p^* \equiv (p - p_{\infty})/(\frac{1}{2}\rho_\infty u_\infty^2)$. The shear layer near the leading edge of the cavity rolls up approximately at $x/D = 2$, generating large spanwise vortices. As large-scale vortices convect downstream, they break into small-scale vortical structures due to the instabilities. These vortical structures eventually impinge on the aft wall, causing large unsteadiness and fluctuations. The pressure wave reflects at the aft wall and travels upstream of the cavity, inducing a feedback mechanism for self-sustained oscillatory flow behavior. In addition, compression waves emanate from the shear layer due to the interaction between large-scale vortices and the freestream, serving as another responsible source of the large pressure fluctuation in the cavity flow, as shown in the numerical schlieren in Fig.~\ref{fig04:ComputationalDomain_3DCF}(b).

\begin{figure*}
    \centering
    \includegraphics[width=1.0\textwidth]{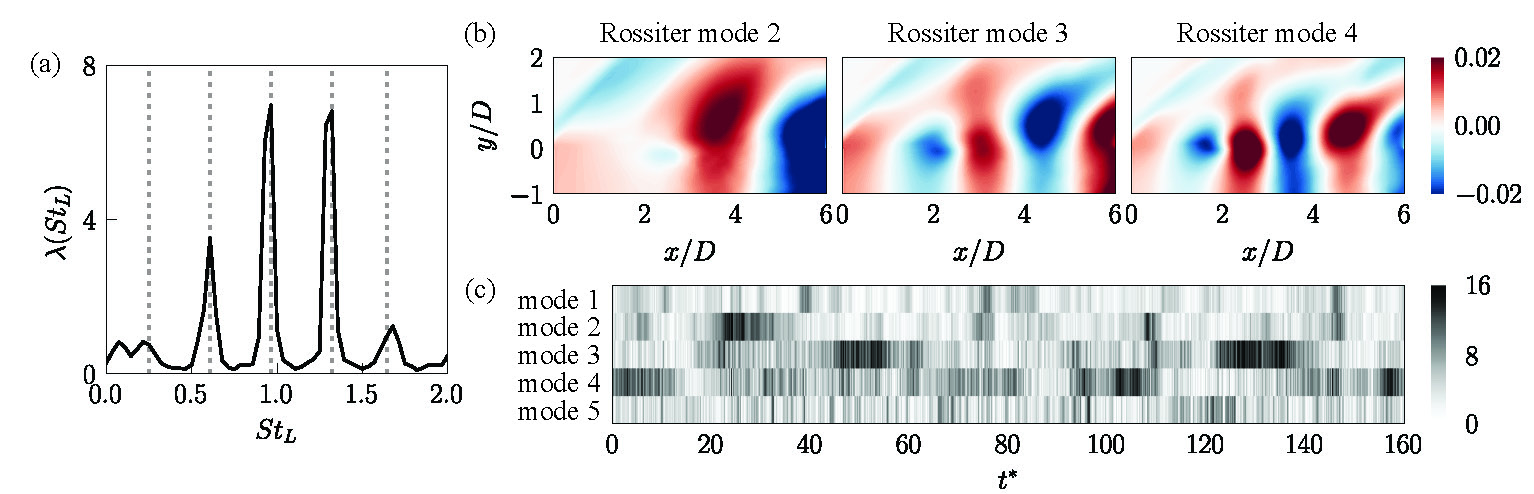}
    \caption{SPOD analysis of the spanwise-averaged pressure field. (a) Frequency spectrum identified by the largest SPOD eigenvalues $\lambda$ (solid) and the Rossiter mode frequencies (dotted), (b) contours of the real part of three dominant SPOD modes and (c) SPOD coefficients of dominant Rossiter modes.
    \label{fig05:SPOD_3DCF}}
\end{figure*}

We further conduct an SPOD analysis of the spanwise-averaged pressure field to identify the dominant frequencies and modal structures of the cavity flow. The SPOD confirms three dominant frequencies at cavity length-based Strouhal numbers of $St_{L} = \Omega L/(2\pi u_{\infty}) \approx 0.61$, 0.96 and 1.32 as shown in the frequency spectrum Fig.~\ref{fig05:SPOD_3DCF}(a), which correspond to the second, third and fourth Rossiter modes. Moreover, we observe that the higher frequencies are associated with small-scale pressure fluctuations in the cavity, as shown in the SPOD modal structures provided in Fig.~\ref{fig05:SPOD_3DCF}(b). To investigate the temporal variation of the strength of the frequency components, we calculate the SPOD coefficients of the first to fifth Rossiter modes by orthogonal projection on their SPOD mode subspace, as shown in Fig.~\ref{fig05:SPOD_3DCF}(c). We observe the switching of dominance among Rossiter modes \citep{kegerise2004mode}, implying that the Rossiter modes exchange and transfer their energy through nonlinear interactions, which makes the oscillator extraction from the flow challenging owing to the intermittently weak frequency components.

\begin{table*}
    \centering
    \caption{\label{tab:Architecture} 
    Neural network architecture of neural networks for oscillator extraction and dynamics modeling.}
% \begin{ruledtabular}
\begin{tabular}{ >{\centering\arraybackslash}m{14em} >{\centering\arraybackslash}m{6em} | >{\centering\arraybackslash}m{14em} >{\centering\arraybackslash}m{6em}}
 \hline \hline
 \multicolumn{4}{c}{Oscillator identifying autoencoder}                                                                                  \\
 \multicolumn{2}{c}{Encoder}                          & \multicolumn{2}{c}{Decoder}                                     \\
 \hline
 Layer                             & Data size        & Layer                             & Data size                   \\
 Pressure field                    & $(192, 96,   1)$ & Oscillator                        & $(      2     )$            \\
 Conv2D $(3, 3,  32)$ + LN + LReLU & $(192, 96,  32)$ & Fully connected + LN + Tanh       & $(     128    )$            \\
 Conv2D $(3, 3,  64)$ + LN + LReLU & $(192, 96,  64)$ & Fully connected + LN + Tanh       & $(    18432   )$            \\
 Max pooling $(2, 2)$              & $( 96, 48,  64)$ & Unflattening                      & $( 12,  6, 256)$            \\
 Conv2D $(3, 3,  64)$ + LN + LReLU & $( 96, 48,  64)$ & Bilinear upsampling $(2, 2)$      & $( 24, 12, 256)$            \\
 Conv2D $(3, 3, 128)$ + LN + LReLU & $( 96, 48, 128)$ & Conv2D $(3, 3, 256)$ + LN + LReLU & $( 24, 12, 256)$            \\
 Max pooling $(2, 2)$              & $( 48, 24, 128)$ & Conv2D $(3, 3, 128)$ + LN + LReLU & $( 24, 12, 128)$            \\
 Conv2D $(3, 3, 128)$ + LN + LReLU & $( 48, 24, 128)$ & Bilinear upsampling $(2, 2)$      & $( 48, 24, 128)$            \\
 Conv2D $(3, 3, 256)$ + LN + LReLU & $( 48, 24, 256)$ & Conv2D $(3, 3, 128)$ + LN + LReLU & $( 48, 24, 128)$            \\
 Max pooling $(2, 2)$              & $( 24, 12, 256)$ & Conv2D $(3, 3,  64)$ + LN + LReLU & $( 48, 24,  64)$            \\
 Conv2D $(3, 3, 256)$ + LN + LReLU & $( 24, 12, 256)$ & Bilinear upsampling $(2, 2)$      & $( 96, 48,  64)$            \\
 Conv2D $(3, 3, 256)$ + LN + LReLU & $( 24, 12, 256)$ & Conv2D $(3, 3,  64)$ + LN + LReLU & $( 96, 48,  64)$            \\
 Max pooling $(2, 2)$              & $( 12,  6, 256)$ & Conv2D $(3, 3,  64)$ + LN + LReLU & $( 96, 48,  64)$            \\
 Flatten                           & $(    18432   )$ & Bilinear upsampling $(2, 2)$      & $(192, 96,  64)$            \\
 Fully connected + LN + Tanh       & $(     128    )$ & Conv2D $(3, 3,  32)$ + LN + LReLU & $(192, 96,  32)$            \\
 Fully connected                   & $(      2     )$ & Conv2D $(3, 3,   1)$              & $(192, 96,   1)$            \\
                                   &    Oscillator    &                                   & Pressure reconstruction     \\
 \hline \hline
 \multicolumn{4}{c}{Neural ODE} \\
 \multicolumn{2}{c}{Dynamics $\boldsymbol{f}$}& \multicolumn{2}{c}{Estimator $\hat{\boldsymbol{\psi}}$}    \\
 \hline
 Layer                  & Data size & Layer                  & Data size                                \\
 Three oscillators      &  $(6) $   & Three oscillators                                                 \\
 Fully connected + Tanh &  $(64)$   & Fully connected + Tanh & $(64)$                                   \\
 Fully connected + Tanh &  $(64)$   & Fully connected + Tanh & $(64)$                                   \\
 Fully connected        &  $(6) $   & Fully connected        & $(8)$                                    \\
 \hline \hline
\end{tabular}
% \end{ruledtabular}
\end{table*}

We now extract the representative oscillators by training oscillator-identifying autoencoders using 12,400 snapshots of the spanwise-averaged pressure field, which covers $t^* = tu_\infty/L \approx 230$. Flow snapshots generated by LES are interpolated to a uniformly discretized domain $(x,y)/D = [0,6] \times [-1,2]$, having points $N_x=192$ and $N_y=96$ in each direction. We split the time series of flow field snapshots into three subsequences with a ratio of 8:1:1 for training, validation and test datasets, respectively. The architecture of the oscillator-identifying autoencoders is summarized in Table.~\ref{tab:Architecture}. LeakyReLU (LReLU) and hyperbolic tangent (Tanh) functions are used for the activation of CNN and MLP, respectively. Layer normalizations (LN) are implemented to reduce the training time of oscillator-identifying autoencoders and improve their generalizability performance \citep{ba2016layer}. The hyperparameters that balance terms in the loss function are carefully tuned through parametric examinations to ensure that the reconstruction performance is not degraded and different frequency components are not introduced into the frequency spectra of individual oscillator variables.

\begin{figure*}
    \centering
    \includegraphics[width=1.0\linewidth]{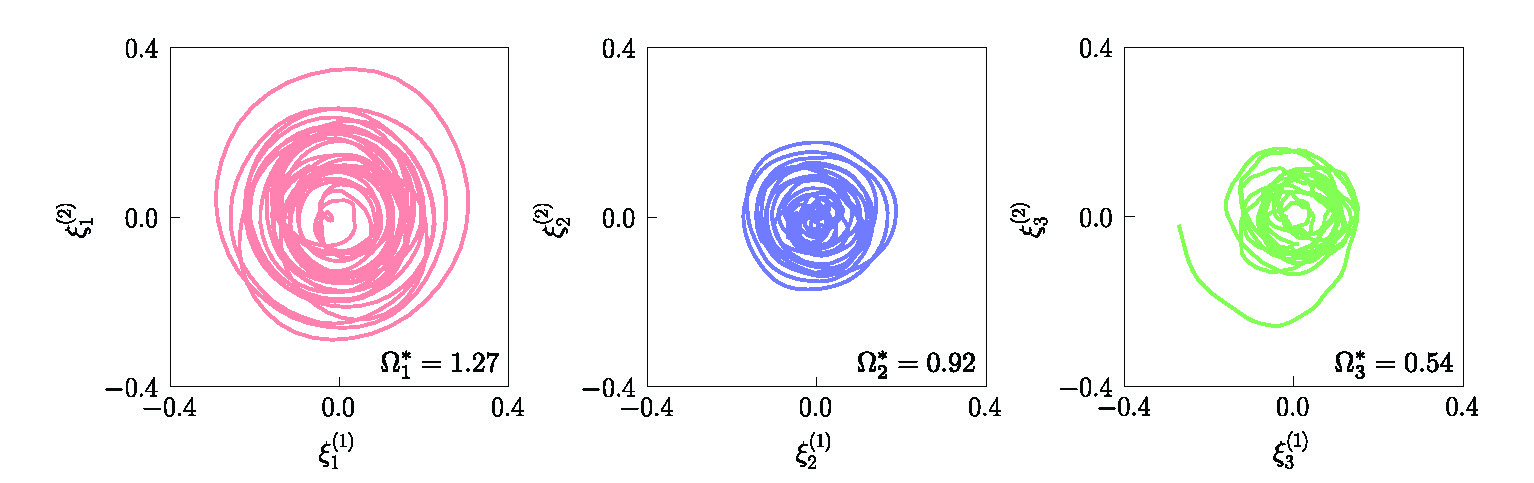}
    \caption{Three representative oscillators extracted by oscillator-identifying autoencoders from the spanwise-averaged pressure field of the cavity flow.
    \label{fig06:Oscillators_3DCF}}
\end{figure*}

\begin{figure*}
    \centering
    \includegraphics[width=1.0\linewidth]{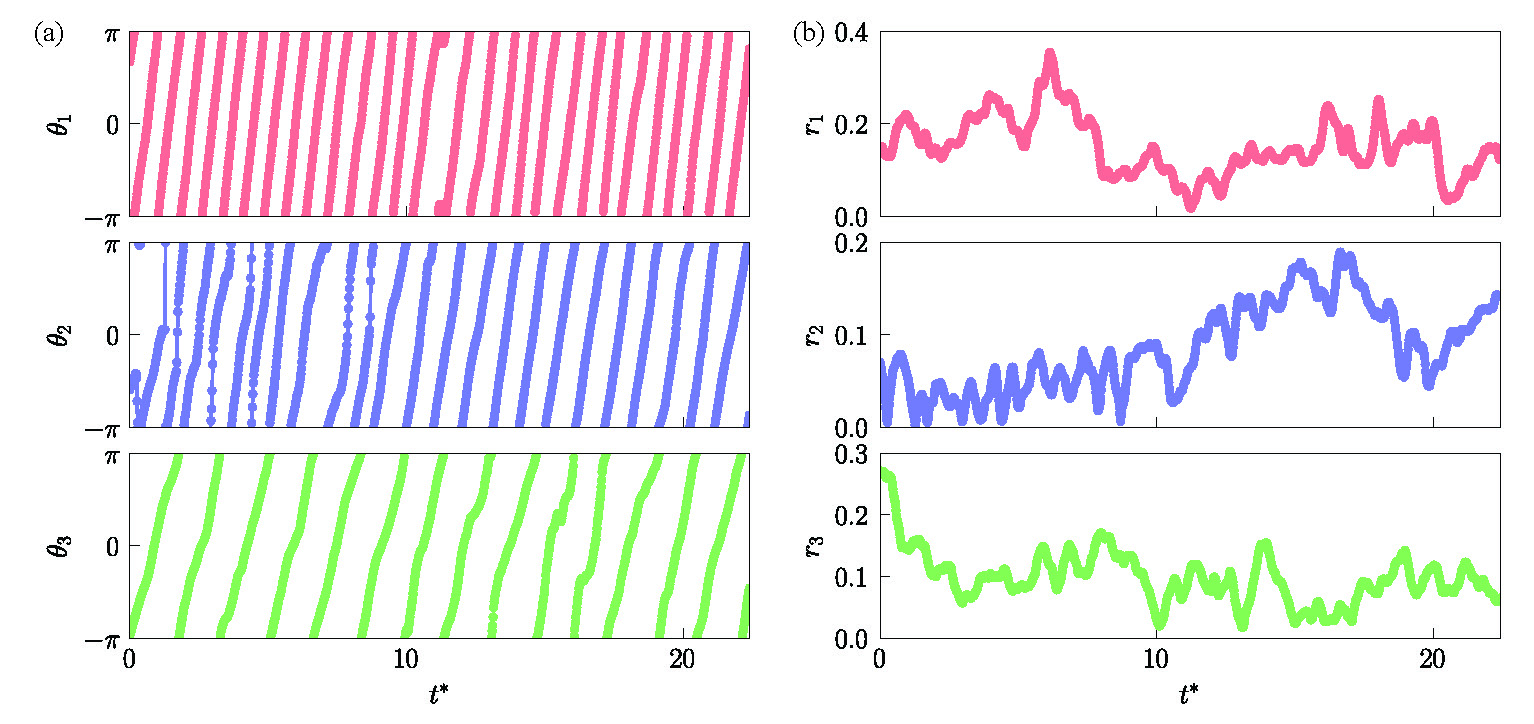}
    \caption{Temporal variation of (a) phase and (b) amplitude variables of oscillators extracted by oscillator-identifying autoencoders.
    \label{fig07:PhaseAmplitude_3DCF}}
\end{figure*}

\begin{figure*}
    \centering
    \includegraphics[width=1.0\linewidth]{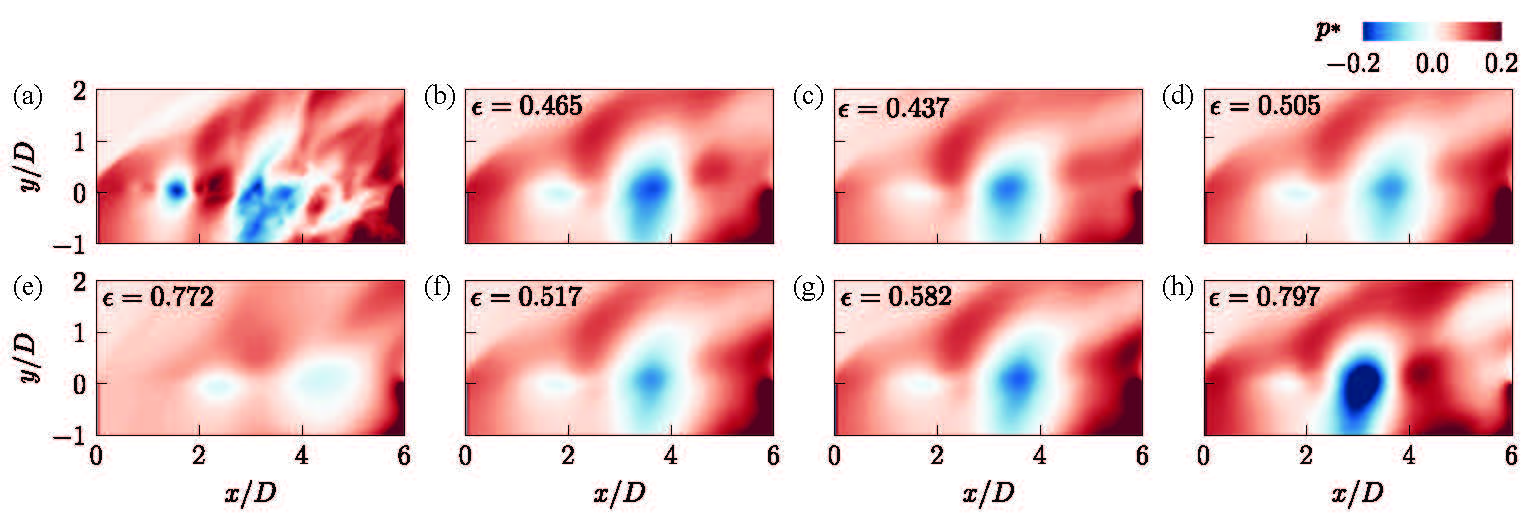}
    \caption{Reconstruction of the spanwise-averaged nondimensionalized pressure field $p^{*}$ by three oscillators at $t^* \approx 18.7$ in the test data and mean relative error $\epsilon = \| p^{*} - \tilde{p}\|/\|p^{*}\|$ based on the Frobenius norm throughout the test dataset. (a) Original pressure field and reconstruction by (b) oscillator-identifying autoencoders and (c) dominant SPOD modes. Neural ODE solution of (d) oscillators extracted by autoencoders and (e) SPOD coefficients without noise ($\varepsilon = 0$) in the observation $\bs{\psi}$. Neural ODE solution of oscillators extracted by autoencoders with the noise level of (f) $\varepsilon = 0.1$, (g) $\varepsilon = 0.3$ and (h) $\varepsilon = 0.5$ in the observation.
    \label{fig08:Reconstruction_3DCF}}
\end{figure*}

Three representative oscillators of cavity flow sequentially extracted by oscillator-identifying autoencoders are shown in Fig.~\ref{fig06:Oscillators_3DCF} based on evaluation with the test data. The extracted oscillators exhibit the frequencies $\Omega_1^* \approx 1.27$, $\Omega_2^* \approx 0.92$ and $\Omega_3^* \approx 0.54$, which match the dominant frequencies identified by the SPOD-based analysis, where the oscillator frequencies are nondimensionalized as $\Omega_m^* = \Omega_m L/(2\pi u_{\infty})$. The extracted oscillators rotate smoothly around the origin while exhibiting amplitude variations over time. The temporal phase and amplitude variations of oscillators are presented in Fig.~\ref{fig07:PhaseAmplitude_3DCF}. We observe that the phase variables consistently vary with the frequency captured by the frequency identifier. The phase variation of the oscillator is less consistent when the amplitude becomes small, since the corresponding oscillatory component is not sufficiently energetic when the oscillator approaches the origin, which reflects the mode switching characteristics of the cavity flow. We also confirm that three representative oscillators can capture large-scale pressure fluctuations in the cavity flow by comparing the reconstructed pressure field with the original pressure field in Fig.~\ref{fig08:Reconstruction_3DCF}(a) and (b), which is quantitatively comparable to the reconstruction by the dominant SPOD modes shown in Fig.~\ref{fig08:Reconstruction_3DCF}(c). These indicate that the physical characteristics and coherent structures in turbulent flows are successfully captured by the oscillators extracted by the proposed method. 

\begin{figure*}
    \centering
    \includegraphics[width=1.0\linewidth]{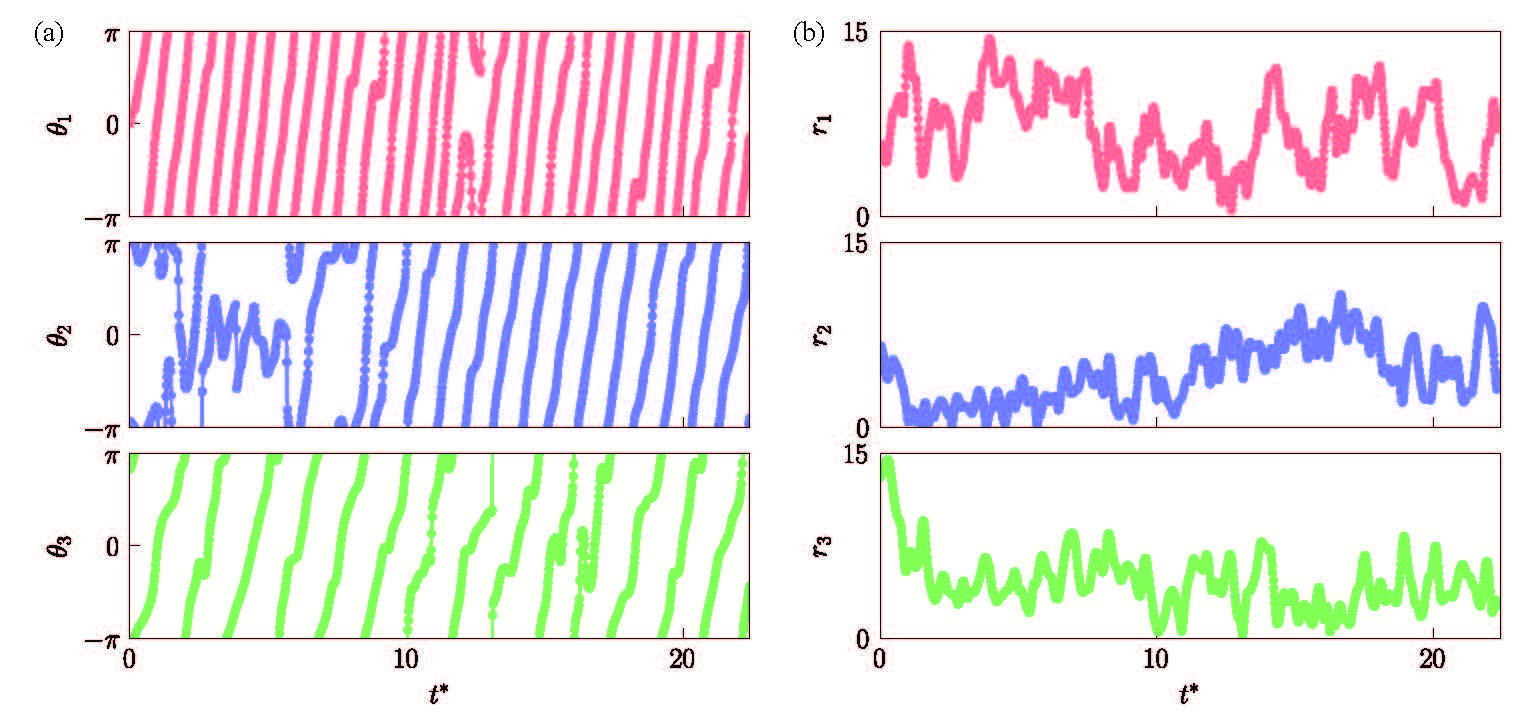}
    \caption{Temporal variation of phase and amplitude variables characterized by the projection onto SPOD modes. 
    \label{fig09:Phase_3DCF-SPOD}}
\end{figure*}

\begin{figure*}
    \centering
    \includegraphics[width=1.0\linewidth]{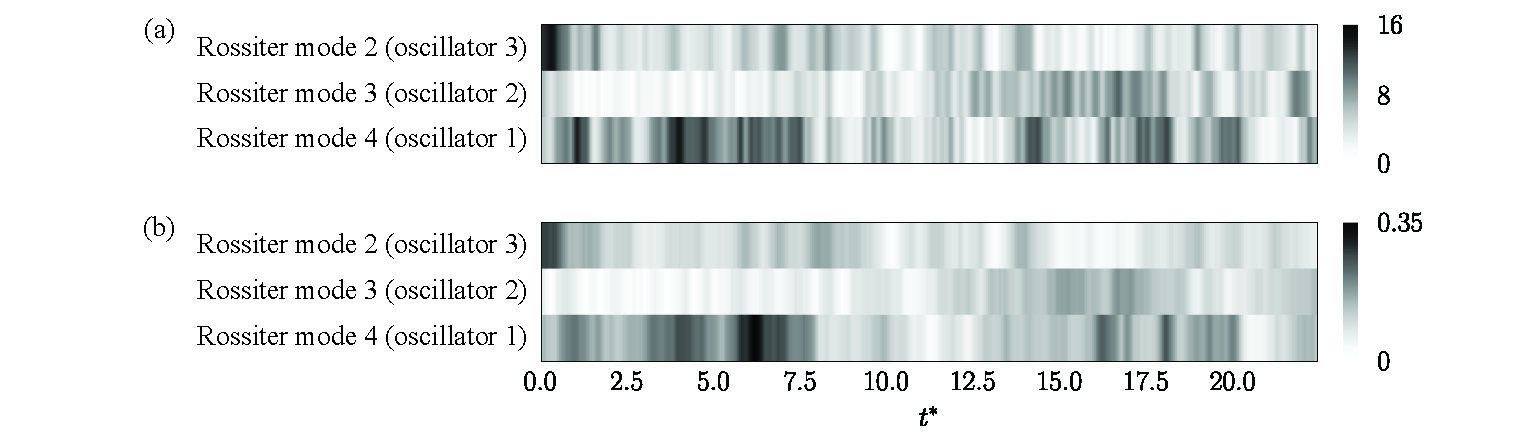}
    \caption{Dominance switching among Rossiter modes in the test data visualized by (a) SPOD coefficients and (b) amplitude variables extracted by oscillator-identifying autoencoders.
    \label{fig09-1:ModeDominanceComparison}}
\end{figure*}

As discussed in Sec.~\ref{sec:2-2}, linear modal analysis techniques identifying the modal structures associated with dominant frequencies, such as DMD and SPOD, have been alternatively employed to define the phase and amplitude variables of turbulent flows. In order to highlight the advantages of the oscillator-identifying autoencoder in oscillator extractions, the temporal variation of phase and amplitude variables defined based on the projection onto SPOD modes is plotted in Fig~\ref{fig09:Phase_3DCF-SPOD}. We observe that the phase variables obtained from SPOD modes lose their consistency in temporal variation when the amplitude variables approach zero. This implies that the oscillator identification based on the linear modal analysis techniques struggles to define phase variables when the frequency components become weak. On the other hand, oscillators extracted by oscillator-identifying autoencoders possess improved consistency in rotation, supported by the nonlinearity and mask function used in training. In addition, the dominance among Rossiter modes in the test data characterized by the SPOD coefficients and amplitude variables obtained from oscillator-identifying autoencoders is compared in Fig.~\ref{fig09-1:ModeDominanceComparison}. While the two approaches share the general trends in the dominance among the modes, the amplitude variable shows smoother temporal amplitude variation over time compared to the SPOD-based method.

Using the oscillators extracted by the oscillator-identifying autoencoders, we construct the data-driven model to infer oscillator dynamics as mentioned in~\ref{sec:2-3}. We use eight instantaneous pressure measurements on the center plane $z = 0$ for the observation $\bs{\psi}$, where four sensors are evenly distributed along the bottom and aft wall of the cavity, respectively. The detailed architecture of the neural networks to model the dynamics function $\bs{f}$ and the estimator $\hat{\bs{\psi}}$ is provided in Table.~\ref{tab:Architecture}. Networks $\bs{f}$ and $\hat{\bs{\psi}}$ are nonlinear MLPs with the hyperbolic tangent function, whereas the gain function $\bs{K}$ is determined as a linear function since it is identified that the nonlinear gain function does not significantly improve the prediction accuracy in the ablation study. Networks $\bs{f}$ and $\bs{K}$ are trained by the derivative-based method and the trajectory-based method, respectively, where the training method for each network is determined after the ablation study. We adopt the 5th-order Dormand–Prince method for the ODE solver, which was verified to be temporally accurate and stable.

\begin{figure*}
    \centering
    \includegraphics[width=1.0\linewidth]{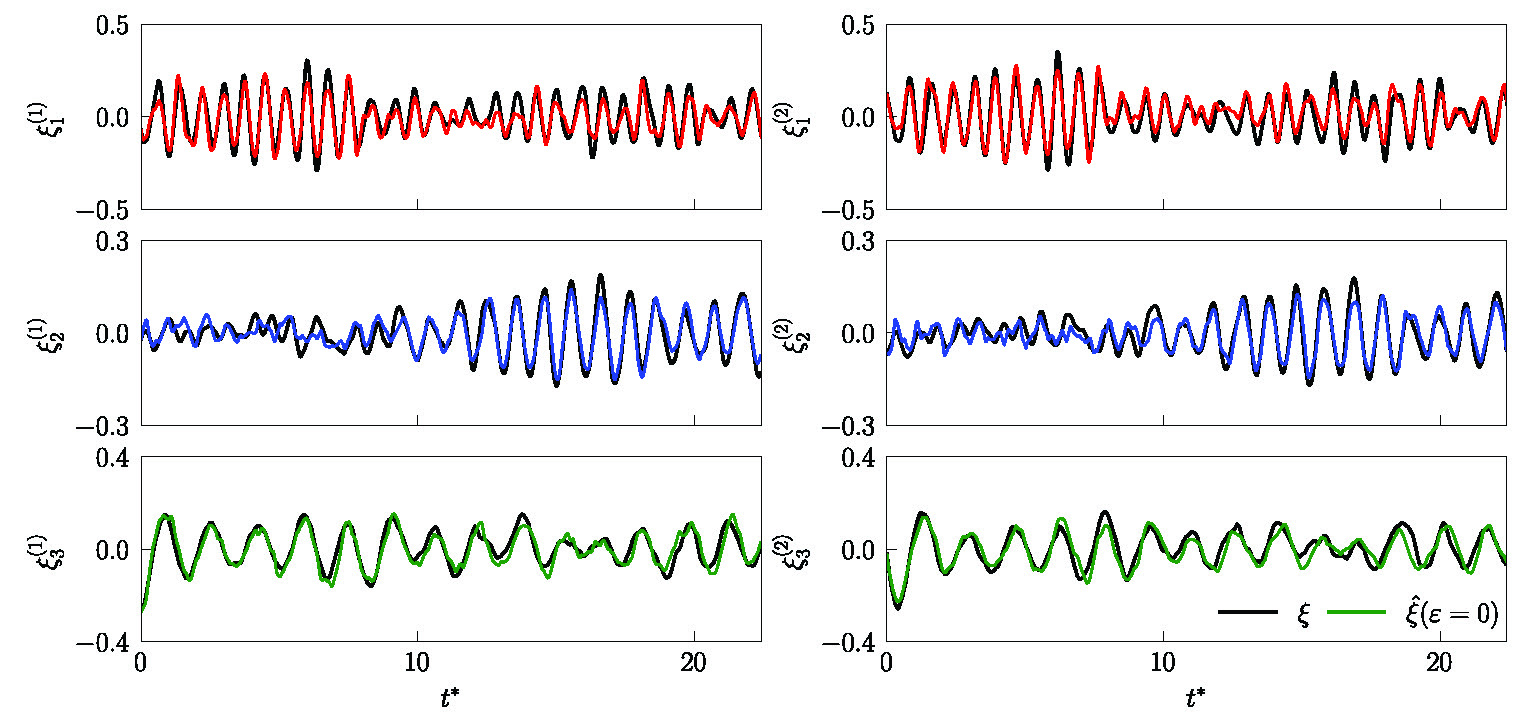}
    \caption{Prediction of the oscillator dynamics of the cavity flow by the neural ODE for the test data.
    \label{fig10:Prediction_NeuralODE}}
\end{figure*}

In order to verify the data-driven oscillator dynamics model, we compare the numerical solution $\hat{\bs{\xi}}(t)$ of Eq.~\ref{eq:LatentDynamics_wCorrection} with the oscillator variable $\bs{\xi}(t)$ provided by oscillator-identifying autoencoders in Fig.~\ref{fig10:Prediction_NeuralODE}. The numerical solution of the trained oscillator dynamics model accurately predicts the oscillator behavior for both short and long prediction horizons, owing to the correction by instantaneously collected information from the pressure sensor measurements. From the ablation study, we confirmed that the prediction largely deviates from the ground-truth oscillator states without the correction term, implying the necessity of the correction term for enabling long-term predictions. While the numerical solution $\hat{\bs{\xi}}(t)$ exhibits smaller errors in the phase prediction, the amplitude shows relatively larger errors. This is because amplitude dynamics are inherently more complicated than phase dynamics, as they reflect energy variation and transfer between different frequency components, contributing to under-resolved amplitude dynamics. The oscillator dynamics model can also reconstruct the pressure field by decoding the numerical solution $\hat{\bs{\xi}}$ using the decoders as shown in Fig.~\ref{fig08:Reconstruction_3DCF}(d). We observe that the decoded neural ODE solution successfully reconstructs the large-scale fluctuations in the pressure field, implying that the developed data-driven dynamics model captures the physical characteristics of the cavity flow.

For further evaluation of the predictive capability of the dynamics model based on the oscillators identified by the autoencoders, we compare it to the equivalent model constructed upon the SPOD-based oscillators. Using the same architecture, we train another neural ODE based upon the SPOD coefficients of three dominant modes to predict the oscillatory behavior of the turbulent cavity flow. The pressure field reconstructed from the linear superposition of the SPOD coefficients predicted by the neural ODE is shown in Fig.~\ref{fig08:Reconstruction_3DCF}(e). Compared to the reconstruction obtained by the model built with the oscillators identified by autoencoders, the neural ODE with SPOD coefficients exhibits larger errors due to inaccurate predictions for the oscillators. This indicates that autoencoders discover more suitable oscillators than projection onto the SPOD modes subspace for modeling and predicting turbulent flow dynamics, attributed to smooth amplitude variables and well-defined phase variables.

\begin{figure*}
    \centering
    \includegraphics[width=1.0\linewidth]{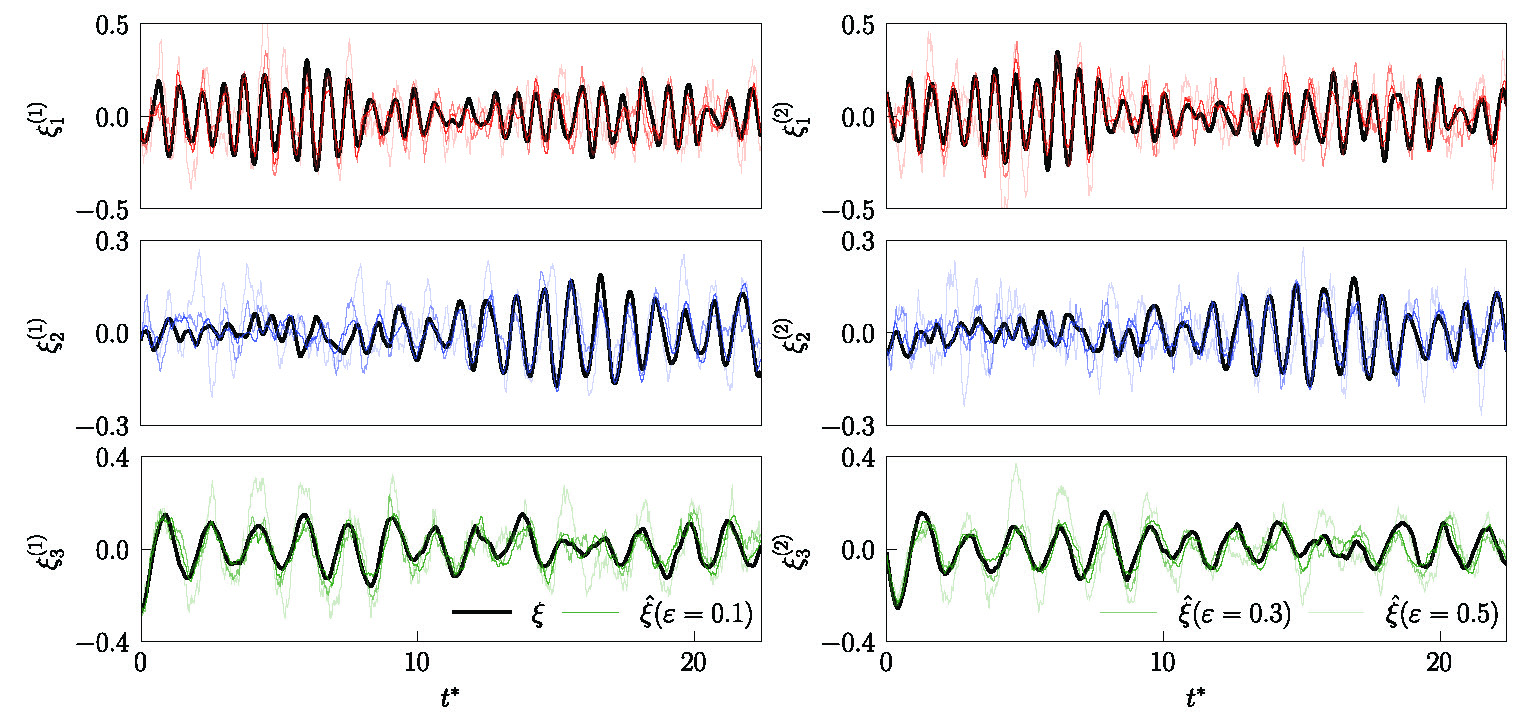}
    \caption{Effect of the noisy observations on the prediction of the oscillator dynamics of the cavity flow.
    \label{fig11:Prediction_NeuralODE_wNoise}}
\end{figure*}

Furthermore, we assess the robustness of the prediction by the present data-driven models to noisy observations. We add a Gaussian noise to the observation to generate the noisy observation $\bs{\psi}^{\mathrm{noise}}$ as
\begin{equation}
    \psi_i^{\mathrm{noise}}(t) = \psi_i(t) + \varepsilon[\max(\psi_i) - \min(\psi_i)]e_i(t),
\end{equation}
where $\psi_i$ is the $i$-th element of $\bs{\psi}$ and $e_i$ is a random variable independently sampled from the standard Gaussian distribution. Three different noise levels are selected as $\varepsilon=0.1, 0.3$ and 0.5 to investigate how $\varepsilon$ affects the accuracy of the prediction model. We confirm that the numerical solution under the noisy observation with up to $\varepsilon = 0.3$ predicts the general trends in the oscillator behavior as shown in Fig.~\ref{fig11:Prediction_NeuralODE_wNoise}. Moreover, the numerical solution of the oscillator dynamics model can still capture the dominant flow structures in the reconstruction as provided in Fig.~\ref{fig08:Reconstruction_3DCF}(f) and (g), even though the observation $\bs{\psi}$ is affected by noise. For $\varepsilon=0.5$, although the amplitude variables are predicted less accurately, the phase variables are still in good agreement with the original oscillator trajectory. As a result, large-scale structures in the reconstruction provided in Fig.~\ref{fig08:Reconstruction_3DCF}(h) share considerable similarity with the reconstruction by oscillator-identifying autoencoders, whereas they obviously exhibit differences in the strength of structures. This demonstrates that the dynamics model developed by the proposed data-driven framework using numerical data has potential to address observations with a significant level of external noise.

\section{Conclusions}    \label{sec:4}

We presented a data-driven approach to extract the phase and amplitude variables from turbulent flows exhibiting multiple dominant frequencies and model their dynamics. We used the parallel structure of autoencoders to reduce the turbulent flow to a set of representative oscillators that reflect its multi-frequency characteristics. Autoencoders were trained with specially designed loss functions accompanied by frequency identifiers, which facilitate the formation of their latent space as oscillators. The dynamics of extracted oscillators were modeled based on the neural ODE, incorporating data assimilation from external observations to predict the behavior of oscillatory components in the turbulent flow.

To demonstrate the proposed data-driven framework, we applied it to the three-dimensional supersonic turbulent cavity flow. From the analysis using SPOD, we identified three dominant Rossiter tones exhibiting switching dominance. We verified that oscillators extracted by the autoencoders precisely capture the dominant frequencies in the flow and reflect the mode-switching characteristics of the cavity flow. Assisted by pressure measurements on cavity walls, the data-driven dynamics model predicted the oscillatory components of the flow with accuracy and robustness against noisy inputs. We also confirmed that the dynamics model with the oscillators identified by the autoencoders shows greater accuracy compared to the equivalent model constructed by the dominant SPOD coefficients.

The data-driven oscillator-based modeling approach proposed in this study not only reduces the complexity of the turbulent flow but also provides us with physical insights into the convection and strength of the oscillatory flow components, which are closely connected to the phase and amplitude variables. Oscillator-identifying autoencoders directly provide phase and amplitude functionals that map the flow field to oscillators, allowing us to gain useful insights for turbulent flow control by computing the sensitivity functions of representative oscillators to external perturbations using automatic differentiation of the encoders. Beyond the theoretical analysis of turbulent flows with multiple dominant frequencies, the reduced-order model enables the real-time prediction of the turbulent flow. Furthermore, the robustness of the dynamics model developed based on numerical data indicates high potential for applications in realistic engineering problems including flow control. 

\section*{Acknowledgments}    \label{sec:acknowledgments}

This work was supported by the Air Force Office of Scientific Research (grant numbers FA9550-22-1-0013 and FA9550-21-1-0178) and the Vannevar Bush Faculty Fellowship (grant number N00014-22-1-2798). The authors are grateful to Chi-An Yeh, Qiong Liu and Vedasri Godavarthi for their advice on numerical settings for flow simulations, and Jonathan Tran for the valuable discussion on data-driven approaches.

\bibliographystyle{unsrtnat}
\bibliography{reference}

\end{document}